\documentclass[a4paper,11pt]{article}
\usepackage[latin1]{inputenc}
\usepackage{amsmath}
\usepackage{graphicx}

\title{\huge Beyond the tree level\\in the AdS/CFT Correspondence}
\author{\Large Mattia Jona-Lasinio\thanks{email : jona@df.unipi.it} \\
\small{Dipartimento di Fisica, Universit\`a di Pisa} \\ 
\small{INFN, Sezione di Pisa} \\
\small{Via Buonarroti 2, 56127 Pisa, Italy}}
\date{February 27, 2002}

\begin{document}

\maketitle

\newcommand{\be}{\begin{equation}}
\newcommand{\ee}{\end{equation}}
\newcommand{\ads}{$AdS_{n+1}$\ }
\newcommand{\fouriexp}{e^{-i\textbf{k}\cdot(\textbf{x}-\textbf{y})}}
\newcommand{\fouriexpr}{e^{-i\textbf{k}\cdot\textbf{r}}}
\newcommand{\misura}[1]{\frac{d^{#1}k}{(2\pi)^#1}\,}
\newcommand{\parg}{\left.\frac{\partial\,G(x,y)}{\partial y_0}\right|_{y_0=\epsilon}}
\renewcommand{\vec}[1]{\overrightarrow{#1}}
\providecommand{\dom}{\partial\Omega}
\providecommand{\sint}[1]{\int_{\dom} d^n{#1}\,}

\vspace{1.5cm}
\begin{abstract}
The loop expansion of the effective action is used to evaluate quantum corrections
to a scalar field theory (massive $\phi^4$ model) on \ads. We evaluate one loop 
corrections and show that they preserve conformal invariance of the boundary theory 
as conjectured by AdS/CFT correspondence.
\end{abstract}
\newpage

\section{Introduction}
It was at the end of 1997 that Maldacena \cite{maldacena-cong} conjectured
a correspondence between the large $N$ limit\footnote{$N$ is the dimension of the gauge group.}
of certain Yang-Mills superconformal field theories and
supergravity on \ads. However in his article he didn't explain how to do calculations with it and
several authors tried to make the correspondence more precise.
In \cite{gubser-klebanov} a possible formulation within string theory can be found, but
it was mainly Witten \cite{witten-ads} that
explained how to construct explicitly the two point Green's function in a general frame.
He considered a simple model like the scalar field theory and using heuristic
arguments, wrote an expression for the Green's function in the massless and massive case.

Basically the idea is to identify the partition function of the bulk (\ads) theory
with the generating functional for the boundary conformal theory. Obviously
the partition function involves a functional integration so we must impose
suitable boundary conditions to restrict this integration. The boundary value
of the bulk field is to be identified as a source term for the boundary theory. In this
formulation the correspondence seems to be true more generally and not restricted to 
string theory. In this paper we consider a massive scalar field theory.

Suppose to have the
$\phi$ field defined on \ads (which we shall denote $\Omega$) and consider the $\phi_0$
field defined as the value
of the $\phi$ field on the boundary $\dom$. At the same time consider a set of operators
$\mathcal{O}$ which belong to the boundary theory and assume a coupling of the form 
$$
\sint{x}\mathcal{O}(x)\phi_0(x)
$$
We can define the generating functional for the boundary theory as
$$
Z[\phi_0]=\left\langle e^{\int_{\dom}\phi_{0}\mathcal{O}}\right\rangle_{CFT}
$$
but we have no idea on how to evaluate the $\langle\rangle_{CFT}$ expectation value.
Consider now the bulk scalar theory and its partition function $Z_{AdS}[S;\phi_0]$ whose 
expression is
$$
Z_{AdS}[S;\phi_0]=\int_{\Omega,\phi_0} \mathcal{D}\phi\,\, e^{-S[\phi]}
$$
The AdS/CFT correspondence states that we can identify the two last expressions
\be \label{relfond}
\left\langle e^{\int_{\dom}\phi_{0}\mathcal{O}}\right\rangle_{CFT} \equiv 
\int_{\Omega,\phi_0} \mathcal{D}\phi\,\, e^{-S[\phi]}
\ee
provided that we restrict the functional integration on the right hand side to
the $\phi$ fields that satisfy the $\phi_0$ boundary condition. In this way
we have a prescription on how to evaluate the $\langle\rangle_{CFT}$ expectation value.

A large number of papers on this subject has been produced starting from \cite{maldacena-cong}
and \cite{witten-ads}. 
See for example \cite{giddings-adscftsmatrix} \nocite{freedman-corrfuncadscft} 
\nocite{arefeva-singletons} \nocite{arutyunov-sugraboundterms} 
- \cite{henningson-spinors} for specific arguments.
A detailed introduction to the AdS/CFT correspondence
can be found in the first part of \cite{petersen-introadscft} and for a general reference
on conformal field theories see \cite{cft}.

To my knowledge the correspondence has always been considered at tree level,
evaluating the partition function of the bulk theory only over the classical solution.
The basic requirement is that the derivatives with respect to $\phi_0$
evaluated at $\phi_0=0$ of the right hand side of \eqref{relfond} are covariant
under the conformal group.
It is then interesting to push further our calculation by considering the one loop
approximation in order to see whether the correspondence remains true. This is what
we do in this paper. We emphasize that our calculations are formal in the sense
that we do not worry about possible infinities in our expressions. Nevertheless infinities
are a real problem which several authors discussed 
(for example \cite{arefeva-cs-breaking}, \cite{muck-renorm1}) but we think
that a formal verification of the transformation properties under the conformal group
is a necessary prerequisite for any further development.
In section 2 we propose a brief
review of the tree level case and the basic results. In section 3 we calculate
the one loop corrections for the correlation functions and we study their transformation
rules.

\section{Scalar field theory on \ads: a review}
In the following we will use some notations introduced in several recent papers 
(\cite{witten-ads}, \cite{petersen-introadscft}).
Consider a (n+2)-dimensional pseudoeuclidean space $E^{n+2}$ with coordinates 
$$
\textbf{Y}\equiv(y_0,\dots,y_{n+1}) \qquad
\qquad ds^2=dy_{0}^2-dy_{n+1}^2-\sum_{i=1}^{n}dy_{i}^2
$$
In this frame the euclidean\footnote{Here the ``time'' coordinate
is represented by $y_{n+1}$, so the euclidean \ads corresponds to considering
$y_{n+1} \rightarrow iy_{n+1}$; the minkowskian version of \ads would have been
$$
y_{0}^2+y_{n+1}^2-\sum_{i=1}^{n}y_{i}^2=1
$$
} \ads equation turns out to be
$$
y_{0}^2-y_{n+1}^2-\sum_{i=1}^{n}y_{i}^2=1
$$
Define $u\equiv y_0+y_{n+1}$, $v\equiv y_0-y_{n+1}$, 
$x_i\equiv\frac{y_i}{u}$ with $i=1,\dots,n$ and $x_0\equiv u^{-1}$. In this way
we obtain the well known \ads metric
$$
ds^2=\frac{1}{x_{0}^2}\sum_{i=0}^{n}dx_{i}^2
$$
The $y$ set of coordinates should not be confused with the $x$ set since the former
is defined in the $E^{n+2}$ space while the latter is defined on the \ads manifold.
The \ads manifold $\Omega$ is represented by the upper half
plane with equation $x_0>0$ while its boundary $\dom$ is represented by the hyperplane 
$x_0=0$ plus the single point $x_0=\infty$.

The action $S[\phi]$ for the scalar $\phi^4$ model is
\be \label{azione}
S[\phi]=\int_{\Omega}
d^{n+1}x\,\sqrt{g}\,\left\{\frac{1}{2}\left[(\partial_{\mu}\phi\partial^{\mu}\phi)+m^2
{\phi}^2\right]+\frac{\lambda}{4!}{\phi}^4\right\}
\ee
which yields the field equation
\be \label{eqmoto}
(\nabla^2-m^2)\phi=\frac{\lambda}{3!}\phi^3
\ee
where\footnote{
In these coordinates the operator $(\nabla^2-m^2)$ is explicitly given by
$$
\left(x_0^2\sum_{i=0}^{n}\frac{\partial ^2}{\partial x_i^2}-x_0(n-1)\frac{\partial}
{\partial x_0}-m^2\right)
$$} $\nabla^2\equiv\frac{1}{\sqrt{g}}\partial_{\mu}(\sqrt{g}\partial^{\mu})$.
Here the $\sqrt{g}$ factor is the square root of the metric determinant and makes
the global measure $\{d^{n+1}x\,\sqrt{g}\}$ invariant under general coordinates
transformations. By means of the Green's formula we transform \eqref{eqmoto} into
an integral equation
\be \label{soluzione}
\begin{split} 
\phi(x)&=\sint{y}\sqrt{h}\,n_\mu\frac{\partial\,G(x,y)}{\partial y_\mu}\phi(y) +
\frac{\lambda}{3!}\int_{\Omega}d^{n+1}y\,\sqrt{g}\,G(x,y)(\phi(y))^3\\
&=\phi^{(0)}(x) + \phi^{(1)}(x)
\end{split}
\ee
Here $h$ is the determinant of the induced metric on $\dom$ and the covariant
Green's function satisfies the equation
$$
(\nabla^2-m^2)G(x,y)=\frac{\delta^{n+1}(x-y)}{\sqrt{g}}
$$
with the boundary condition $\left. G(x,y)\right|_{x \in \dom}=0$.  
The tree level approximation is then $\phi(x)\simeq\phi^{(0)}(x)$ and
the action takes the form
\be \label{azionecl}
S[\phi]=\frac{1}{2}\sint{x}\sqrt{h}\,n_{\mu}\phi^{(0)}\partial^{\mu}\phi^{(0)} + 
\frac{\lambda}{4!}\int_{\Omega}d^{n+1}x\,\sqrt{g}\,(\phi^{(0)})^4
\ee

Let's consider now the free field case.
Solving equation \eqref{eqmoto} by Fourier-transform methods, we find 
two linearly independent solutions
$$
x_0^{\frac{n}{2}}e^{-i\textbf{k}\cdot\textbf{x}}I_{\alpha}(kx_0) \qquad
x_0^{\frac{n}{2}}e^{-i\textbf{k}\cdot\textbf{x}}K_{\alpha}(kx_0)
$$
where $I_{\alpha}$ e $K_{\alpha}$ are modified Bessel
functions\footnote{$I_{\alpha}(z)$ and $K_{\alpha}(z)$ satisfy the differential
equation
$$
z^2\frac{d^2w}{dz^2}+z\frac{dw}{dz}-(z^2+\alpha^2)w=0
$$
}
\cite{abram} and $\alpha=\sqrt{\frac{n^2}{4}+m^2}$; $\textbf{k}$ is an n dimensional
vector while $k=|\textbf{k}|\equiv \sqrt{\sum_{i=1}^{n}k_i^2}$. The covariant Green's function is
\begin{multline} \label{greenf}
G(x,y)=-(x_0y_0)^{\frac{n}{2}} \int \misura{n}\fouriexp\,
\left[K_\alpha(kx_0)\,I_\alpha(ky_0)\,\theta(x_0-y_0)\, + \right.\\ \left.
+\,I_\alpha(kx_0)\,K_\alpha(ky_0)\,\theta(y_0-x_0)\right]
\end{multline}
where $\theta$ is the Heaviside step function. This expression can be integrated
explicitly and the result is
\be \label{greenint}
G(x,y)=-\frac{c}{2\alpha}\xi^{-\Delta}F\left(\frac{n}{2},\Delta;\alpha+1;\xi^{-2}\right)
\ee
where
$$
\Delta=\frac{n}{2}+\alpha \qquad
c=\frac{\Gamma(\Delta)}{\pi^{\frac{n}{2}}\Gamma(\alpha)}
$$
and $F(a,b;c;z)$ is the hypergeometric function; the expression for $\xi$ is
$$
\xi=\frac{1}{2x_0y_0}\left\{\frac{1}{2}\left[|x-y|^2+|x-y^*|^2\right]+\sqrt{|x-y|^2
|x-y^*|^2}\right\}
$$
where $x\equiv(x_0,\textbf{x})$, $x^*\equiv(-x_0,\textbf{x})$ and $|u|^2 \equiv 
\sum_{i=0}^{n} u_i^2$.

After a rather lengthy calculation it is possible to
arrive at the following expression for the action $S[\phi]$ at tree level
\be
\begin{split} \label{azionetree}
S[\phi]=&-c\alpha\int_{\dom}d^{n}x\,d^{n}y\,
\frac{\phi_{0}(x)\phi_{0}(y)}{|\textbf{x}-\textbf{y}|^{2\Delta}}+\\
&+\frac{\lambda \, c^4}{4!}\int_{\dom}d^{n}x_1\dots d^{n}x_4\,\, \phi_0(x_1) \dots
\phi_0(x_4)\, I_4(x_1,\dots,x_4)
\end{split}
\ee
where $I_{4}(x_1,\dots,x_4)$ is expressed by
\be
I_4(x_1,\dots,x_4)=\int_{\Omega}\frac{d^{n+1}y}{y_0^{n+1}}
\left(\frac{y_0}{y_0^2+|\textbf{y}-\textbf{x}_1|^2}\right)^{\Delta}
\dots
\left(\frac{y_0}{y_0^2+|\textbf{y}-\textbf{x}_4|^2}\right)^{\Delta}
\ee
and $\phi_{0}$ represents the boundary value of the $\phi$ field.
The AdS/CFT correspondence states then that two and four point functions
for the boundary theory must have the form
\be \label{twopointtree}
\langle\mathcal{O}(\textbf{x})\mathcal{O}(\textbf{y})\rangle \propto \frac{1}
{|\textbf{x}-\textbf{y}|^{2\Delta}}
\ee
\be \label{fourpointtree}
\langle\mathcal{O}(\textbf{x}_1)\mathcal{O}(\textbf{x}_2)\mathcal{O}(\textbf{x}_3)\mathcal{O}(\textbf{x}_4)
\rangle \propto f\left(\frac{\textbf{x}_{12}\textbf{x}_{34}}{\textbf{x}_{13}\textbf{x}_{24}},
\frac{\textbf{x}_{12}\textbf{x}_{34}}{\textbf{x}_{14}\textbf{x}_{23}}\right)
\prod_{i,j=1\,;\,i<j}^{4}\;\frac{1}{\textbf{x}_{ij}^{\frac{2}{3}\Delta}}
\ee
as dictated by conformal invariance. In \cite{muck-ads/cft} 
it is shown that this is the case and the argument is
generalized to the n point functions at tree level. Here $f$ is an arbitrary function and
$\textbf{x}_{ij} \equiv |\textbf{x}_i - \textbf{x}_j|$.

\section{Beyond the tree level: one loop corrections}
In the following we discuss in detail the two and four point one loop corrections and we show that
the argument can be extended to the n point case straightforwardly.

\subsection{Effective action}
Let's define $\phi_{J}(x)$ to be the classical solution, which obeys the classical
field equation $\left.\frac{\delta S}{\delta \phi
(x)}\right|_{\phi=\phi_{J}}\equiv J(x)$ and satisfies $\phi_{J}(x)|_{J=0}=0$.
Define the fluctuation field $\xi\equiv \phi - \phi_{J}$ and perform a Taylor
expansion of $S[\phi]$ around the stationary point $\phi=\phi_{J}$. Denoting
the generating functional by $Z[J]$ we have
\be \label{zapprox}
Z[J]\simeq e^{-S[\phi_{J}]+\int J\phi_{J}} 
\left(\frac{\int \mathcal{D}\xi \,e^{-\frac{1}{2}\int d^{n+1}x\, d^{n+1}y \,\sqrt{g(x)}\,\sqrt{g(y)}\,
\left.\frac{\delta^2S}{\delta\xi(x)
\delta\xi(y)}\right|_{\xi=0}\xi(x)\xi(y)}}
{\left.\int \mathcal{D}\xi \,e^{-\frac{1}{2}\int d^{n+1}x\, d^{n+1}y \,\sqrt{g(x)}\,\sqrt{g(y)}\,
\left.\frac{\delta^2S}{\delta\xi(x)
\delta\xi(y)}\right|_{\xi=0}\xi(x)\xi(y)}\right|_{J=0}}\right)
\ee
Note that $\left.\frac{\delta^2S}{\delta\xi(x)\delta\xi(y)}\right|_{\xi=0}=
\left.\frac{\delta^2S}{\delta\phi(x)\delta\phi(y)}\right|_{\phi=\phi_{J}}$ and the
integrals involved in \eqref{zapprox} are Gaussian, so we have 
\be
Z[J]\simeq e^{-S[\phi_{J}]+\int J\phi_{J}}\frac{\sqrt{\det \left(\left.\frac{\delta^2S}
{\delta\phi(x)\delta\phi(y)}\right|_{\phi=0}\right)}}{\sqrt{\det \left(\left.\frac{\delta^2S}
{\delta\phi(x)\delta\phi(y)}\right|_{\phi=\phi_{J}}\right)}}
\ee
The effective action $\Gamma$ is
\be
\Gamma[\phi_{J}]=S[\phi_{J}] + \frac{1}{2} \ln \det \left(\frac{\left.\frac{\delta^2S}
{\delta\phi(x)\delta\phi(y)}\right|_{\phi=\phi_{J}}}{\left.\frac{\delta^2S}
{\delta\phi(x)\delta\phi(y)}\right|_{\phi=0}}\right)=S[\phi_{J}] + \frac{1}{2} \ln \det 
\left(A[\phi^{(0)}]\right)
\ee
where $\phi(J;x)\equiv\frac{1}{\sqrt{g(x)}}\frac{\delta \,ln \,Z[J]}{\delta J(x)}$ and
we have defined
\be
A[\phi^{(0)}]\equiv \frac{x_0^2\sum_{i=0}^{n}\frac{\partial ^2}{\partial x_i^2}-x_0(n-1)
\frac{\partial}{\partial x_0}-m^2-\frac{\lambda}{2}(\phi^{(0)}(x))^2}{x_0^2\sum_{i=0}^{n}
\frac{\partial ^2}{\partial x_i^2}-x_0(n-1) \frac{\partial}{\partial x_0}-m^2}
\ee

\subsection{One loop corrections}
The one loop corrections for the two and four point functions are
\be \label{twopoint}
\langle\mathcal{O}(\textbf{y})\mathcal{O}(\textbf{z})\rangle=
\frac{a\lambda c^2}{2} \,\int \frac{d^{n+1}x}{x_0^{n+1}}\,
\left(\frac{x_0}{x_0^2+|\textbf{x}-\textbf{y}|^2}\right)^\Delta
\left(\frac{x_0}{x_0^2+|\textbf{x}-\textbf{z}|^2}\right)^\Delta
\ee
\begin{multline} \label{fourpoint}
\langle\mathcal{O}(\textbf{y})\mathcal{O}(\textbf{z})\mathcal{O}(\textbf{v})
\mathcal{O}(\textbf{w})\rangle
=\frac{3}{2}\,\lambda^2 c^4
\int\frac{d^{n+1}x}{x_0^{n+1}}\frac{d^{n+1}x^{\prime}}{x_0^{\prime
n+1}}G^2(x,x^{\prime})\times \\
\times 
\left(\frac{x_0}{x_0^2+|\textbf{x}-\textbf{y}|^2}\right)^\Delta
\left(\frac{x_0}{x_0^2+|\textbf{x}-\textbf{z}|^2}\right)^\Delta
\left(\frac{x_0^{\prime}}{x_0^{\prime2}+|\textbf{x$^{\prime}$}-\textbf{v}|^2}\right)^\Delta
\left(\frac{x_0^{\prime}}{x_0^{\prime2}+|\textbf{x$^{\prime}$}-\textbf{w}|^2}\right)^\Delta
\end{multline}
We have here defined $a \equiv G(x,x)$ because it is easy to check that $G(x,x)$ is a constant. 

We now show that these expressions satisfy the constraints imposed by conformal invariance.
It is clear from the very structure that they are invariant under the Poincar\'e
subgroup, so all we have to do is to verify their covariance under dilations and SCT (Special
Conformal Transformations).

Under a dilation $\textbf{u} \rightarrow \alpha \textbf{u}$ we have
\be
\langle\mathcal{O}(\alpha\textbf{y})\mathcal{O}(\alpha\textbf{z})\rangle=
\frac{a\lambda c^2}{2} \,\int \frac{d^{n+1}x}{x_0^{n+1}}\,
\left(\frac{x_0}{x_0^2+|\textbf{x}-\alpha\textbf{y}|^2}\right)^\Delta
\left(\frac{x_0}{x_0^2+|\textbf{x}-\alpha\textbf{z}|^2}\right)^\Delta
\ee
Changing the integration variable $x \rightarrow \alpha x$ we have
\be
\begin{split}
\langle\mathcal{O}(\alpha\textbf{y})\mathcal{O}(\alpha\textbf{z})\rangle&=
\frac{a\lambda c^2}{2} \,\frac{\alpha^{2\Delta}}{\alpha^{4\Delta}}\,\int 
\frac{d^{n+1}x}{x_0^{n+1}}\,
\left(\frac{x_0}{x_0^2+|\textbf{x}-\textbf{y}|^2}\right)^\Delta
\left(\frac{x_0}{x_0^2+|\textbf{x}-\textbf{z}|^2}\right)^\Delta\\
&=\alpha^{-2\Delta}\langle\mathcal{O}(\textbf{y})\mathcal{O}(\textbf{z})\rangle
\end{split}
\ee
The dilation is performed on an n-vector while the change in the integration
variable is on an (n+1)-vector, but we can always interpret $\textbf{u}
\equiv(0,\textbf{u})$.

Let's consider now the SCT transformation $\textbf{u} \rightarrow \textbf{u}^{\prime}=
\frac{\textbf{u}-\textbf{b}\textbf{u}^2}{1-2(\textbf{b}\cdot\textbf{u})+
\textbf{b}^2\textbf{u}^2}$ and define
$$
\gamma(\textbf{u})\equiv [1-2(\textbf{b}\cdot\textbf{u})+\textbf{b}^2\textbf{u}^2]
$$
$$
\gamma(u)\equiv [1-2(b \cdot u)+ b^2 u^2]
$$
where the dot stands for the canonical scalar product.

We have
$$
\langle\mathcal{O}(\textbf{y}^{\prime})\mathcal{O}(\textbf{z}^{\prime})\rangle=
\frac{a\lambda c^2}{2} \,\int \frac{d^{n+1}x}{x_0^{n+1}}\,
\left(\frac{x_0}{x_0^2+\left|\textbf{x}-\frac{\textbf{y}-\textbf{b}\textbf{y}^2}
{\gamma(\textbf{y})}\right|^2}\right)^\Delta
\left(\frac{x_0}{x_0^2+\left|\textbf{x}-\frac{\textbf{z}-\textbf{b}\textbf{z}^2}
{\gamma(\textbf{z})}\right|^2}\right)^\Delta
$$
Note that $\gamma(u) \neq \gamma(\textbf{u})$ so that the $\gamma$ factors cannot be
reabsorbed by changing the integration variable.
In this case the verification requires three preliminary remarks:
\begin{itemize}
	\item an n dimensional SCT transformation with parameter $\textbf{b}$ performed
	on n dimensional vectors $\textbf{u}$ is equivalent to an n+1 dimensional SCT
	transformation with parameter $b=(0,\textbf{b})$ performed on n+1 dimensional
	vectors $u=(0,\textbf{u})$;
	\item an n+1 dimensional SCT transformation with parameter $\textbf{b}$ performed
	on n+1 dimensional vectors $u=(u_0,\textbf{u})$ is explicitly given by
	$$
	\begin{cases}
	u_0 \rightarrow \frac{u_0}{\gamma(u)}\\
	\textbf{u} \rightarrow \frac{\textbf{u}-\textbf{b}\textbf{u}^2}{\gamma(u)}
	\end{cases}
	$$
	so that the ``time'' component is only rescaled;
	\item equation \eqref{twopoint} can be written as
	\be \label{twopoint1}
	\langle\mathcal{O}(\textbf{y})\mathcal{O}(\textbf{z})\rangle=
	\frac{a\lambda c^2}{2} \,\int \frac{d^{n+1}x}{x_0^{n+1}}\,
	\left(\frac{x_0}{|x-y|^2}\right)^\Delta
	\left(\frac{x_0}{|x-z|^2}\right)^\Delta
	\ee
	where $y=(0,\textbf{y})$, $z=(0,\textbf{z})$.
\end{itemize}
Consider then the n-dimensional transformation with parameter \textbf{b}
as a particular n+1 dimensional transformation with parameter $b=(0,\textbf{b})$.
We shall note that under SCT we have
$$
|x_i-x_j| \rightarrow \frac{|x_i-x_j|}{\sqrt{\gamma(x_i)\gamma(x_j)}}
$$
so that equation \eqref{twopoint1} transforms into
\be
\begin{split}
\langle\mathcal{O}(\textbf{y}^{\prime})\mathcal{O}(\textbf{z}^{\prime})\rangle&=
\frac{a\lambda c^2}{2} \,\int \frac{d^{n+1}x}{x_0^{n+1}}\,
\left(\frac{x_0\gamma(x)\gamma(y)}{\gamma(x)|x-y|^2}\right)^\Delta
\left(\frac{x_0\gamma(x)\gamma(z)}{\gamma(x)|x-z|^2}\right)^\Delta\\
&=\gamma(y)^{\Delta}\gamma(z)^{\Delta}
\langle\mathcal{O}(\textbf{y})\mathcal{O}(\textbf{z})\rangle
\end{split}
\ee
but since we have assumed $b=(0,\textbf{b})$ and $u=(0,\textbf{u})$,
we have $\gamma(u)=\gamma(\textbf{u})$ whence
\be
\langle\mathcal{O}(\textbf{y}^{\prime})\mathcal{O}(\textbf{z}^{\prime})\rangle=
\gamma(\textbf{y})^{\Delta}\gamma(\textbf{z})^{\Delta}
\langle\mathcal{O}(\textbf{y})\mathcal{O}(\textbf{z})\rangle
\ee
accordingly to what was expected.

The four point case parallels the two point one.
It is straightforward to
show covariance under dilations and special conformal transformations:
the argument is exactly the same. The only difference is the Green's
function $G^2(x,x^{\prime})$ which is now not constant but we show that 
in fact it is an invariant function. We remind its expression
$$
G(x,x^{\prime})=-\frac{c}{2\alpha}\xi^{-\Delta}F\left(\frac{n}{2},\Delta;\alpha+1;\xi^{-2}\right)
$$
where
$$
\xi=\frac{1}{2x_0x^{\prime}_0}\left\{\frac{1}{2}\left[|x-x^{\prime}|^2+|x-x^{\prime\,*}|^2\right]
+\sqrt{|x-x^{\prime}|^2 |x-x^{\prime\,*}|^2}\right\}
$$
and $x\equiv(x_0,\textbf{x})$, $x^*\equiv(-x_0,\textbf{x})$.

Dilation invariance of the $\xi$ variable is self evident.
Under SCT, $\xi$ transforms into
$$
\xi=\frac{\gamma(x)\gamma(x^{\prime})}{2x_0x^{\prime}_0}\left\{\frac{1}{2}\left[
\frac{|x-x^{\prime}|^2}{\gamma(x)\gamma(x^{\prime})}+\frac{
|x-x^{\prime\,*}|^2}{\gamma(x)\gamma(x^{\prime\,*})}\right]+\frac{\sqrt{|x-x^{\prime}|^2
|x-x^{\prime\,*}|^2}}{\gamma(x)\sqrt{\gamma(x^{\prime})\gamma(x^{\prime\,*})}}\right\}
$$
but since $b=(0,\textbf{b})$, we have $\gamma(u)=\gamma(u^{*})$.
We can conclude that the variable $\xi$ is invariant under SCT too and write
\be
\langle\mathcal{O}(\alpha\textbf{y})\mathcal{O}(\alpha\textbf{z})
\mathcal{O}(\alpha\textbf{v})\mathcal{O}(\alpha\textbf{w})\rangle=
\alpha^{-4\Delta}\langle\mathcal{O}(\textbf{y})\mathcal{O}(\textbf{z})
\mathcal{O}(\textbf{v})\mathcal{O}(\textbf{w})\rangle
\ee
\be
\langle\mathcal{O}(\textbf{y}^{\prime})\mathcal{O}(\textbf{z}^{\prime})
\mathcal{O}(\textbf{v}^{\prime})\mathcal{O}(\textbf{w}^{\prime})\rangle=
\gamma(\textbf{y})^{\Delta}\gamma(\textbf{z})^{\Delta}
\gamma(\textbf{v})^{\Delta}\gamma(\textbf{w})^{\Delta}
\langle\mathcal{O}(\textbf{y})\mathcal{O}(\textbf{z})
\mathcal{O}(\textbf{v})\mathcal{O}(\textbf{w})\rangle
\ee

To generalize the argument it is not difficult to write down the 2N point
function\footnote{The number of points must be even because of the
$\phi \rightarrow -\phi$ symmetry in the action. All functions with an odd 
number of points vanish.}, following the hint given by the two and four point case. We have
\begin{multline}
\langle\mathcal{O}(\textbf{x}_{1a})\mathcal{O}(\textbf{x}_{1b}) \dots 
\mathcal{O}(\textbf{x}_{Na})\mathcal{O}(\textbf{x}_{Nb})\rangle
\propto \\
\propto \lambda^{N} \, c^{2N}\int \frac{d^{n+1}z_{1}}{{z_1}_0^{n+1}} \dots \frac{d^{n+1}z_{N}}{{z_N}_0^{n+1}}\, \,
G(z_1,z_2) \, G(z_2,z_3) \dots G(z_{N-1},z_N) \, G(z_N,z_1) \times \\
\times \left(\frac{{z_1}_0}{{z_1}_0^2+|\textbf{z}_1-\textbf{x}_{1a}|^2}\right)^\Delta
\left(\frac{{z_1}_0}{{z_1}_0^2+|\textbf{z}_1-\textbf{x}_{1b}|^2}\right)^\Delta \times
\dots \\
\dots \times \left(\frac{{z_N}_0}{{z_N}_0^2+|\textbf{z}_N-\textbf{x}_{Na}|^2}\right)^\Delta
\left(\frac{{z_N}_0}{{z_N}_0^2+|\textbf{z}_N-\textbf{x}_{Nb}|^2}\right)^\Delta
\end{multline}
Every step of our argument can be replicated, so we can say that also at one loop
level the AdS/CFT correspondence holds
in the sense that one loop corrections have the same transformation
laws under the conformal group as the correlation functions at tree level.


\section*{Acknowledgments}
I am grateful to Prof. Kensuke Yoshida for introducing me to the AdS/CFT correspondence.

\bibliographystyle{mybib}
\bibliography{bibliografia}

\end{document}